\documentclass{article} 
\usepackage{times}
\RequirePackage[colorlinks,citecolor=blue,urlcolor=blue]{hyperref}
\usepackage{url}
\usepackage{npbayes}
\usepackage{amsmath,amsthm,amssymb,bm,graphicx,url,comment}
\allowdisplaybreaks

\newcommand{\commentt}[1]{}

\newcommand{\lam}{\boldsymbol{\Lambda}}

\title{Variational Bayes for Merging Noisy Databases}

\author{
Tamara Broderick\\
University of California, Berkeley\\
\texttt{tab@stat.berkeley.edu}
\and
Rebecca C.~Steorts \\
Carnegie Mellon University \\
\texttt{beka@cmu.edu}
}

\begin{document}

\maketitle

\begin{abstract}
Bayesian entity resolution merges together multiple, noisy databases
and returns the minimal collection of unique individuals represented, together
with their true, latent record values. 
Bayesian methods allow flexible generative models
that share power across
databases as well as principled quantification of uncertainty for queries of
the final, resolved database. However, existing Bayesian methods for
entity resolution use Markov monte Carlo method (MCMC) approximations and are too slow
to run on modern databases containing millions or billions of records. Instead, we propose 
applying variational approximations to allow scalable Bayesian
inference in these models. We derive a coordinate-ascent approximation
for mean-field variational Bayes, qualitatively
compare our algorithm to existing methods, note unique challenges for inference that arise
from the expected distribution of cluster sizes in entity resolution, and discuss directions for future work in this domain.
\end{abstract}

\section{Introduction}

Merging records from multiple databases is a problem that emerged from the genetics literature \cite{newcombe_1959} and is a pressing issue in statistics and computer science in the modern day \cite{christen_2011}. 
For instance, human rights organizations collect records of war crimes in the Middle East and Central America
and want to estimate the total number of victims \cite{lum_2013}. The United States Census Bureau
wants to estimate minority representation and child poverty in different parts of
the country \cite{datta_2011, bell_2013}.
In each of these examples, individual records are collected in multiple databases.
Due to the collection procedure,
records are often duplicated within a single database and across databases.
Crucially, due to various factors, some records in these databases
are corrupted by noise. In any case, an important part of delivering an estimate
for any quantity of interest from the merged databases is also 
returning some uncertainty for that estimate.
\cite{steorts_2014_eb, steorts_2014_aistats, steorts_2014_smered, sadinle_2014} have
recently applied a Bayesian statistical paradigm
to merging databases by modeling the noisy corruption as a random
process; the authors have shown that their approach provides not
only desirable uncertainty quantification for a variety of
model queries but also flexible generative models
to capture the many unique types of records and record relationships that may
be present in these databases. However, the MCMC approximations used in
these Bayesian analyses do not scale sufficiently to process the large number
of records in many modern and complex databases.
Thus, we propose a variational Bayes approximation to capture
desired uncertainty in posterior Bayesian estimates while simultaneously
allowing the processing of much larger and more realistic databases than is possible with these methods. Finally, we elaborate on how these database-merging models pose unique challenges for variational approximations.


\section{Background} \label{sec:background}

\emph{Entity resolution} refers to the merging of multiple
databases (often without shared unique identifiers) into a single database of unique entities \cite{christen_2011}.
Special cases of entity resolution include \emph{record linkage},
which refers to the identification of records across different
databases that represent the same entity, and
\emph{de-duplication}, which refers to the identification of
records within the same database that represent the same entity.
Traditional
approaches for entity resolution
that link records directly to other records become computationally
infeasible as the number of records grows \cite{christen_2011,winkler_2006}.
Here, we instead take the
approach of \cite{steorts_2014_smered}
and imagine each record as representing a latent individual.
When we further suppose that some entries may suffer from noisy
corruption, entity resolution can be viewed
as a clustering problem. The observed data being clustered are the records
in each database, and the latent cluster centers are the unobserved,
latent individuals.


It is common for record fields to be discrete or categorical: e.g.,
county of residence, race, gender, etc. Like \cite{steorts_2014_aistats},
we focus on categorical data in what follows.\footnote{The nature of
data corruption in the database collection process can often lead to
interesting and nonstandard noise distributions for data that might, in other
contexts, be treated as non-categorical or continuous (date of birth,
age, etc.). Text fields---e.g., name of
an individual---must also be treated with more care than assigning a
single categorical distribution in this context. These
considerations, though addressed elsewhere \cite{sadinle_2014, steorts_2014_eb},
are outside the scope of this note.}
\cite{blei_2003_latent, blei_2006_variational} previously demonstrated the
scaling advantages of variational Bayesian approximations
to posteriors for mixture and admixture modeling, where
the observed data are categorically-valued (e.g.\ words in a vocabulary). There exist unique
challenges in our data clustering problem. A particularly important
one lies in an assumption inherent in many of the popular Bayesian
models for clustering and admixture---such as mixture models, LDA \cite{blei_2003_latent}, 
Dirichlet processes \cite{maceachern_1998_estimating},
hierarchical Dirichlet processes \cite{teh_2006_hierarchical}, and many more.
These models all implicitly assume that any cluster makes up a 
non-zero proportion of the data that does not change as the data set size increases 
without bound \cite{kingman_1978_representation, broderick_2013_feature}.
This assumption, by contrast, is
very clearly inappropriate for entity resolution problems. In entity resolution,
we expect each cluster to contain perhaps one and at most a handful
of records. We expect the number of clusters to grow linearly with the 
number of records in a given database---though the number of clusters might be constant
as more databases are added. We call the problem of modeling this
type of clustering behavior, which differs from classical models
and assumptions, the 
\emph{small clustering problem}, and we discuss it in more detail in \mysec{compare}.

\section{A new generative model for entity resolution} \label{sec:gen_model}

Let $D$ represent the number of databases and $R_{d}$ represents
the number of records in the $d$th database. All records contain the 
same $F$ fields, which are categorical. Let field $f$ have $V_{f}$ possible values,
or \emph{field attributes}.
Assume for now that every record is complete;
that is, for every database and record, there is no missing field.
Let $x_{drf}$ be the observed data value in the $f$th field of the $r$th record
in the $d$th database. We make the further simplifying assumption
 that there are $K$ unique latent individuals. Ultimately, we
desire a model where $K$ is random, and we learn a posterior distribution over
$K$ given the full data $x = \{x_{drf}\}_{d,r,f}$ across \emph{all} databases, records,
and fields. But as a first step we assume $K$ is fixed and known, as in LDA \cite{blei_2003_latent}.
Let $z_{dr}$ be the latent individual for whom (potentially noisy) data
$x_{dr\cdot}$ is recorded in the $r$th record of the $d$th database.

In other words, we regard each record $x_{dr\cdot}$
as a possibly distorted copy of an ideal latent record for latent individual $z_{dr}$.
To capture this idea, let $\beta_{kf\cdot}$ be a discrete noise distribution associated with the $k$th
latent individual.
That is, $\beta_{kfv},\,v\in\{1,\ldots,V_f\},$ are numbers between zero and one
that sum to one across $v$. 
If there were no noise in the data entry procedure, the probabilities~$\beta_{kf\cdot}$
would correspond to a trivial distribution with all of its mass at some true latent value $v^{*}_{kf}$
of the $f$th field for the $k$th individual:
$\beta_{kfv} = \mbo\{v = v^{*}_{kf}\}.\footnote{Here,
$\mbo(E)$ is the indicator function for event $E$.}$
In general, there is some noise in the records, and $\beta_{kfv}$ corresponds to a non-trivial noise distribution; however we assume it has a plurality
of its mass at the true value.


For our generative model, assume that the observed value~$x_{drf}$
of the $f$th field in record $r$ in database~$d$ is drawn from the
noise distribution associated with the latent individual~$z_{dr}$ for this record; 
that is,
$$
	x_{drf} | \beta_{\cdot f \cdot}, z_{dr} \sim \cat_{V_{f}}(\beta_{z_{dr} f \cdot}),
$$
where $\cat_{V_{f}}$ is the categorical distribution over $1,\ldots,V_{f}$ with
probabilities given by the distribution parameter. These draws are
independent across records and fields, conditional on $\beta$ and~$z$.

Next, we form a hierarchical Bayesian  model by putting priors on both $z$
and $\beta$. For $z$, we assume that the latent individual for any record
is drawn uniformly over all latent individuals and independently across records:
$$
	z_{dr} \sim \cat_{K}(K^{-1}\,\bfo_{K}),
$$
where $\bfo_{K}$ is the vector of all ones of length $K$.
Since $\beta_{kf\cdot}$ is a vector of probabilities, a natural choice of
prior for $\beta_{kf\cdot}$ is the Dirichlet distribution on a vector of size $V_{f}$, which
we denote by
$\dir_{V_{f}}$.  Thus, we assume that the $\beta_{kf\cdot}$ vectors are drawn independently according to
$$
	(\beta_{kfv})_{v=1}^{V_{f}} \sim \dir_{V_{f}}(A_{\cdot}),
$$
with hyperparameter vector $A_{\cdot} = (A_{1},\ldots,A_{V_{f}})$.
Typically, we assume that the $A_v$ are small (near zero) so that the Dirichlet parameter
encourages $\beta_{kf\cdot}$ to be peaked around a single value.
We typically choose
$A_{1} = \cdots = A_{V_{f}}$.

\section{Comparison with previous work} \label{sec:compare}



We briefly review the model of \cite{steorts_2014_smered}, where the authors introduced
the basic Bayesian clustering framework for entity resolution and their Split and MErge REcord linkage and De-duplication (SMERED) algorithm. 
\cite{steorts_2014_smered} took a fully hierarchical-Bayesian approach, in the
special case where all the record fields are categorical and independent. 
The authors derived an efficient
hybrid (Metropolis-within-Gibbs) MCMC algorithm, SMERED.  SMERED is able to
update most of the latent variables and parameters using Gibbs sampling steps from
conjugate conditional distributions.  While SMERED updates the assignment of
records to latent individuals using a
split-merge step, following \cite{jain_2004}, and can run on a health care databases of 60,000 records in 3.5 hours, it does not scale to ``large databases."\footnote{This database is the National Long Term Care Study (NLTCS), a longitudinal study of the health status of elderly Americans \url{http://www.nltcs.aas.duke.edu/}. The authors ran the NLTCS on three databases of 20,000 records each, for 1 million iterations of their hybrid MCMC, which took 3.5 hours to run.} In terms of scalability, we wish to scale to millions or billions of records in one or multiple databases. For example, the U.S. Census contains approximately 300 million records, while many medical databases at large universities or in the entire country would contain millions or billions of records.


Furthermore, while the model of \cite{steorts_2014_smered} was shown to work very well for entity resolution applications, it is not easily approximated with variational methods due to various deterministic dependencies in the 
generative model. We show the full model for SMERED in Appendix \ref{app:notation}, where we also provide a mapping between the SMERED model and our new generative model from \mysec{gen_model}. By contrast, we directly demonstrate in \mysec{mf_approx} how our new generative model, which is inspired by LDA, is readily amenable to variational approximation.

While our model has some similarities to LDA, there are also some differences---large and small. For one, the fields do not enjoy the symmetry of words in a bag-of-words model of a document; that is, the fields are ordered and cannot be interchanged. Second, as we do not expect the distribution of individuals to vary wildly by database, we keep the same uniform distribution over latent individuals in each database. By contrast, an important part of LDA is allowing the admixture proportions of topics to vary by document.

We now raise a key issue regarding the main distinction between many 
classical Bayesian models for mixtures and admixtures (such as 
mixture models,
LDA, feature-allocation models \cite{broderick_2013_feature} including the
Indian buffet process \cite{griffiths_2005_infinite},
etc.)
and entity resolution. 
The issue arises from framing entity resolution as a clustering problem.
In clustering and other statistical models,
it is common to assume that our data are \emph{infinitely exchangeable}, meaning that 
for any data set size, we assume that the distribution
of our data would not change if the data were observed in a different order.
This simple assumption applied to clustering models implies,
via the \emph{Kingman paintbox}
\cite{kingman_1978_representation,broderick_2013_feature},
that every cluster forms some strictly positive proportion of the data,
and this proportion does not change as the data grows.
In mixture models, these are the mixing proportions; there may be finitely many
in a finite mixture model or infinitely many in a Dirichlet process model.
In any of these cases, there are two important consequences for
our model. First, as the data set size grows, we always
observe more data points in a cluster. In fact, the number of observed data 
in a cluster grows without bound.
Second, because the size of every cluster grows to infinity
as the data set size grows to infinity, the usual
asymptotic theory applies to inferring cluster properties or parameters.
Uncertainty about, e.g., a cluster mean typically shrinks to zero in the limit.

When clusters are unique individuals in a population, however, it is not
natural to assume that more data always eventually means more records of the
same individual.  Rather, every cluster should be observed a strictly finite
number of times.  This means that uncertainty about latent individuals cannot
shrink to zero (in general).  
Since the assumptions of the traditional models (such as LDA) are violated
in this case, they do not apply. And we must ask: what are  natural regularity
assumptions in this \emph{small clustering} domain, what inferences
can we draw about clusters in this domain, and what new families of distributions
can we apply?
A similar issue to what we have dubbed the small clustering problem
has previously been identified for infinitely exchangeable
graphs by \cite{lloyd_2012_random}.
This problem is also reminiscent of challenges in
high-dimensional statistics, where the number of parameters
may grow linearly (or much faster) than the data size.

\section{Mean-field variational approximation} \label{sec:mf_approx}

The generative model specified in \mysec{gen_model}
yields the following joint distribution for data $x$ and parameters $\beta, z$:
\begin{equation} \label{eq:p_joint}
	p(\beta, z, x) =
		\left[
			\prod_{k=1}^{K} \prod_{f=1}^{F} \dir_{V_{f}}(\beta_{kf \cdot} | A_\cdot)
		\right]
		\left[ 
			\prod_{d=1}^{D} \prod_{r=1}^{R_{d}} \prod_{f=1}^{F} \beta_{z_{dr} l x_{drf}}
		\right].
\end{equation}
Note that the posterior on the parameters, $p(\beta, z|x)$, is proportional to $p(\beta, z, x)$.

As this posterior cannot be solved for in closed form, we must approximate it. Here,
we consider a variational approximation $q$ of the following form:
\begin{equation} \label{eq:q_joint}
	q(\beta, z) =
		\left[
			\prod_{d=1}^{D} \prod_{r=1}^{R_{d}} q(z_{dr} | \phi_{dr\cdot})
		\right]
		\left[
			\prod_{k=1}^{K} \prod_{f=1}^{F} q(\beta_{kf\cdot} | \lambda_{kf\cdot})
		\right],
\end{equation}
where we have introduced variational
parameters $\phi$ and $\lambda$. We further assume
\begin{equation} \label{eq:q_indiv}
	q(z_{dr} | \phi_{dr\cdot})
		= \cat_{K}(\phi_{dr\cdot})
	\quad \textrm{and} \quad
	q(\beta_{kf\cdot} | \lambda_{kf\cdot})
		= \dir_{V_{f}}(\beta_{kf\cdot} | \lambda_{kf\cdot}).
\end{equation}

The variational optimization problem is to minimize 
the Kullback-Leibler divergence from $q(\beta,z)$ to $p(\beta,z|x)$:
$
	\min_{\lambda,\phi} \kl\left(q_{\lambda,\phi}(\beta, z)
	\vphantom{\textstyle\frac00}
	\,\middle\|\, p(\beta,z|x)\right).
$
To clarify that the optimization is over different choices of the distribution
$q$, which are indexed by parameters $\lambda$ and~$\phi$,
we write $q(\beta,z)$ as $q_{\lambda,\phi}(\beta,z)$ above.
We derive the following coordinate-ascent steps in the variational parameters $\phi$ and $\lambda$
for the variational optimization problem
$\min_{\lambda,\phi} \kl(q_{\lambda,\phi} \| p)$
in \app{deriv}:
\begin{align*}
	\lambda_{kfv}
		&\leftarrow A_{v} + \sum_{d=1}^{D} \sum_{r=1}^{R_{d}} \phi_{drk} \mbo\{x_{drf} = v\}, \\
	\phi_{drk}
		&\propto_{k} \exp\left\{ 
			\sum_{f=1}^{F} \sum_{v=1}^{V_{f}} 
				\mbo\{x_{drf} = v\}
				\left[ \psi( \lambda_{kfv} )
					- \psi\left( \sum_{u=1}^{V_{f}} \lambda_{kfv} \right) \right]
		\right\}.
\end{align*}

\section{Future directions} \label{sec:future}
Our next step is to compare the algorithm resulting from our new generative model
and variational approximation to the existing Bayesian model and resulting MCMC  algorithm
SMERED of \cite{steorts_2014_smered}. We anticipate the variational approach will be much faster, however, there may be accuracy tradeoffs. Also, since 
we have made many simplifying
assumptions, we propose incorporation of more realistic assumptions about record fields as in \cite{sadinle_2014, steorts_2014_eb}. Moreover,
it remains to allow the number of latent individuals to grow with the size of the data, to construct a model
that allows posterior inference of this number, and to address the small clustering problem we have posed. We wish to find a solution that addresses this problem not only for entity resolution but more broadly in other domains, where clusters may not be expected to grow without bound as a proportion of the total data.

\section{Acknowledgements}

TB was supported by the Berkeley Fellowship.
RCS was supported by NSF grants SES1130706 and DMS1043903 and NIH grant \#1 U24 GM110707-01.

\newpage
\bibliography{vb_er} 
\bibliographystyle{plain}

\newpage
\appendix

\section{Review of SMERED and notational map with new generative model}
 \label{app:notation}

The independent fields model of \cite{steorts_2014_smered} assumes
the $d$ databases are conditionally independent, given the latent
individuals, and that fields are independent within individuals. We use the same notation 
as the generative model earlier, with $D$ databases, $R_{d}$ records within the $d$th database, and $F$ fields within each record. Then $\bm{x}_{dr}$ is a categorical vector of length $p$.  Let $\bm{y}_{k}$
be the latent vector of true field
values
for the $k$th record, where $k \in \{1,\ldots,K\}$ indexes the latent individuals.
The \emph{linkage structure} is defined as
$\bm{\Lambda}=\{\lambda_{dr}\;;\;d=1,\ldots,D\;;\;r=1,\ldots,R_{d}\}$ where
$\lambda_{dr}$ is an integer from $1$ to $K$ indicating the latent
individual to which the $r$th record in database $d$ refers, i.e., $\bm{x}_{dr}$ is a
possibly-distorted measurement of $\bm{y}_{\lambda_{dr}}$.
Finally,
$\tilde{z}_{drf}$ is $1$ or $0$ according to whether or not the particular
field $f$ is distorted in $\bm{x}_{dr}.$

The Bayesian parametric model is
%
\begin{align}
\bm{x}_{drf}\mid\lambda_{dr},\bm{y}_{\lambda_{dr}\ell},z_{drf},\bm{\theta}_f
	&\stackrel{\text{ind}}{\sim}
		\begin{cases}
		\delta_{\bm{y}_{\lambda_{dr} f }}&\text{ if }z_{drf}=0\\
		\cat(1,\bm{\theta}_f)&\text{ if }z_{drf}=1
		\end{cases}\\
	\nonumber
	\tilde{z}_{drf}
		&\stackrel{\text{ind}}{\sim}\bern(\tilde{\beta}_f)\\
	\bm{y}_{kf}\mid\bm{\theta}_{f}
		&\stackrel{\text{ind}}{\sim}\cat(1,\bm{\theta}_f)\\
	\bm{\theta}_f
		&\stackrel{\text{ind}}{\sim}\dir(\bm{\mu}_f)\\
	\tilde{\beta}_f
		&\stackrel{\text{ind}}{\sim}\tb(a_f,b_f) \\
	\pi(\lam)
		&\propto 1.
\end{align}

To compare the SMERED generative model to our generative model in \mysec{gen_model},
first note that the observed $r$th record in database $d$ is $x_{dr\cdot}$ in both models.
The latent individual for the record at $(d,r)$ in SMERED is $\lambda_{dr}$ and in \mysec{gen_model}
it is $z_{dr}$.
In SMERED, the noise distribution for a latent individual is separated into two steps:
whether there is noise for a given field ($\tilde{z}_{drf}$) and the distribution of that noise
($\bm{\theta}_{f}$). By contrast, in \mysec{gen_model}, $\beta_{kf\cdot}$ captures the full
distribution of field values for individual $k$.
Also, while SMERED places a distribution $\bm{\theta}_{f}$ on the underlying distribution
of field values, such a distribution is implicit in aggregating over $\beta_{kf\cdot}$ in
\mysec{gen_model}. Likewise, the ``true record values'' of SMERED's $\bm{y}_{\lambda_{dr} \cdot}$
are implicit in the distribution $\beta_{z_{dr}f\cdot}$ of \mysec{gen_model}.

\section{Mean-field variational approximation derivation} \label{app:deriv}

\subsection{Mean-field variational problem}

We recall that minimizing the Kullback Leibler divergence (KL) divergence,
$$
	\min_{\lambda,\phi} \kl(q_{\lambda,\phi}(\beta, z) || p(\beta,z|x)),
$$
is equivalent to maximizing $\elbo(\phi, \lambda)$, where
\begin{align*}
	\elbo(\phi, \lambda)
		&= -\kl(q_{\lambda,\phi}(\beta, z) || p(\beta,z|x)) + p(x) \\
		&= \mbe_{q}[ \log p(\beta,z,x) ] - \mbe_{q}[ \log q(\beta, z; \phi, \lambda) ].
\end{align*}
Henceforth, we concentrate on maximizing $\elbo(\phi, \lambda)$ with 
respect to mean-field approximation parameters $\phi, \lambda$.

From the generative model in \mysec{gen_model}, 
we derived the joint distribution of parameters $\beta,z$
and data $x$, $p(\beta,z,x)$, in \eq{p_joint}.
We also assume that the approximating distribution $q(\beta,z)$ 
for the posterior $p(\beta, z |x)$ takes the form specified
in \eqs{q_joint} and \eqss{q_indiv}. Using these equations, we find
\begin{align*}
	\elbo(\phi, \lambda)
		&= \sum_{k=1}^{K} \sum_{f=1}^{F} \mbe_{q}[ \log \dir_{V_{f}}(\beta_{kf\cdot} | A) ] \\
			& {} + \sum_{d=1}^{D} \sum_{r=1}^{R_{d}} \sum_{f=1}^{F} \sum_{k=1}^{K} \sum_{v=1}^{V_{f}} \mbe_{q}[ \mbo\{z_{dr} = k\} \mbo\{x_{drf} = v\} \log( \beta_{k f v} ) ] \\
			& {} - \sum_{d=1}^{D} \sum_{r=1}^{R_{d}} \sum_{k=1}^{K}  \mbe_{q}[ \mbo\{z_{dr} = k\} \log ( \phi_{dr k} ) ] \\
			& {} - \sum_{k=1}^{K} \sum_{f=1}^{F} \mbe_{q}[ \log \dir_{V_{f}}(\beta_{kf\cdot} | \lambda_{kf\cdot}) ].
\end{align*}
To evaluate these expectations, we recall the definitions of the \emph{digamma}
$\psi$ and \emph{trigamma} functions $\psi_{1}$:
\begin{align*}
	\psi(x) &= \frac{d }{d x} \log \Gamma(x) \\
	\psi_{1}(x) &= \frac{d^{2}}{dx^{2}} \log \Gamma(x) = \frac{d}{dx} \psi(x).
\end{align*}
With these functions in hand, we can write
\begin{align*}
	\lefteqn{
		\mbe_{q}[ \log \dir_{V_{f}}(\beta_{kf\cdot} | A) ]
		} \\
		&\quad = \log \Gamma(\sum_{v=1}^{V_{f}} A_{v})
			- \sum_{v=1}^{V_{f}} \log \Gamma( A_{v} )
			+ \sum_{v=1}^{V_{f}} (A_{v} - 1) \left[ \psi( \lambda_{kfv} )
				- \psi( \sum_{u=1}^{V_{f}} \lambda_{kfu} ) \right] \\
	\lefteqn{
		\mbe_{q}[ \mbo\{z_{dr} = k\} \mbo\{x_{drf} = v\} \log( \beta_{k f v} ) ]
		} \\
		&\quad = \phi_{drk} \mbo\{x_{drf} = v\}
			\left[ \psi( \lambda_{kfv} )
				- \psi( \sum_{u=1}^{V_{f}} \lambda_{kfv} ) \right] \\
	\lefteqn{
		\mbe_{q}[ \mbo\{z_{dr} = k\} \log ( \phi_{dr k} ) ] 
		} \\
		&\quad = \phi_{drk} \log ( \phi_{dr k} ) \\
	\lefteqn{
		\mbe_{q}[ \log \dir_{V_{f}}(\beta_{kf\cdot} | \lambda_{kf\cdot}) ]
		} \\
		&\quad = \log \Gamma(\sum_{v=1}^{V_{f}} \lambda_{kfv})
			- \sum_{v=1}^{V_{f}} \log \Gamma( \lambda_{kfv} )
			+ \sum_{v=1}^{V_{f}} (\lambda_{kfv} - 1) \left[ \psi( \lambda_{kfv} )
				- \psi( \sum_{u=1}^{V_{f}} \lambda_{kfu} ) \right].
\end{align*}

\subsection{Coordinate ascent}
We find a local maximum of the ELBO via coordinate ascent in each dimension of the variational
parameters: $\lambda, \phi$. This method is sometimes known as \emph{batch} variational inference.

First we look at $\lambda$; the partial derivative of the ELBO with respect to $\lambda_{kfv}$ is
\begin{align*}
	\frac{\partial}{\partial \lambda_{kfv}} \elbo(\phi, \lambda)
		&= (A_{v}-1) \psi_{1}(\lambda_{kfv})
				+ \left[ \sum_{u=1}^{V_{f}} (A_{u} - 1) \right] \psi_{1}( \sum_{u=1}^{V_{f}} \lambda_{kfu} ) \\
			& {} + \sum_{d=1}^{D} \sum_{r=1}^{R_{d}} \phi_{drk} \mbo\{x_{drf} = v\}
				\left[ \psi_{1}(\lambda_{kfv}) - \psi_{1}( \sum_{u=1}^{V_{f}} \lambda_{kfu} ) \right] \\
			& {} + \sum_{u: u \ne v} \sum_{d=1}^{D} \sum_{r=1}^{R_{d}} \phi_{drk} \mbo\{x_{drf} = u\}
				\left[ - \psi_{1}( \sum_{t=1}^{V_{f}} \lambda_{kft} ) \right] \\
			& {} - \psi( \sum_{u=1}^{V_{f}} \lambda_{kfu} ) + \psi(\lambda_{kfv}) \\
			& {} - \left\{ (\lambda_{kfv} - 1) 
				\cdot \left[ \psi_{1}(\lambda_{kfv}) - \psi_{1}( \sum_{u=1}^{V_{f}} \lambda_{kfu} ) \right] \right\} \\
			& {} - \left[ \psi(\lambda_{kfv}) - \psi( \sum_{u=1}^{V_{f}} \lambda_{kfu} ) \right] \\
			& {} - \sum_{u: u \ne v} (\lambda_{kfu} - 1) \psi_{1}( \sum_{t=1}^{V_{f}} \lambda_{kft} ) \\
		&= \psi_{1}(\lambda_{kfv})
				\left[ A_{v} - \lambda_{kfv}
				+ \sum_{d=1}^{D} \sum_{r=1}^{R_{d}} \phi_{drk} \mbo\{x_{drf} = v\} \right] \\
			& {} - \psi_{1}(  \sum_{u=1}^{V_{f}} \lambda_{kfu} )
				\sum_{u=1}^{V_{f}}
				\left[	A_{u} - \lambda_{kfu} + \sum_{d=1}^{D} \sum_{r=1}^{R_{d}} \phi_{drk} \mbo\{x_{drf} = u\} \right]
\end{align*}
This quantity will be zero for
$$
	\lambda_{kfv} \leftarrow A_{v} + \sum_{d=1}^{D} \sum_{r=1}^{R_{d}} \phi_{drk} \mbo\{x_{drf} = v\}.
$$

Next we consider $\phi$.
The partial derivative of the ELBO with respect to $\phi_{drk}$ is
\begin{align*}
	\frac{\partial}{\partial \phi_{drk}} \elbo(\phi, \lambda)
		&= \sum_{f=1}^{F} \sum_{v=1}^{V_{f}} 
			\mbo\{x_{drf} = v\}
			\left[ \psi( \lambda_{kfv} )
				- \psi( \sum_{u=1}^{V_{f}} \lambda_{kfu} ) \right] \\
			& {} - \log(\phi_{drk}) - 1.
\end{align*}
This quantity will be zero, and the $\phi_{drk}$ will sum across $k$ to one,\footnote{This
derivation can be completed
using the Lagrange method of multipliers.} if
$$
	\phi_{drk} \propto_{k} \exp\left\{ 
			\sum_{f=1}^{F} \sum_{v=1}^{V_{f}} 
				\mbo\{x_{drf} = v\}
				\left[ \psi( \lambda_{kfv} )
					- \psi( \sum_{u=1}^{V_{f}} \lambda_{kfu} ) \right]
		\right\}.
$$
We note that the $\phi_{drk}$ satisfy the constraint $\phi_{drk} > 0$ for all $k$
and so form a proper probability distribution across $k$.

\end{document}